\documentclass[aps,12pt,amsmath,amssymb,superscriptaddress,showpacs,showkeys]{revtex4}

\usepackage{dcolumn}
\usepackage{graphicx}
\usepackage{textcomp}
\begin {document}
  \newcommand {\nc} {\newcommand}
  \nc {\beq} {\begin{eqnarray}}
  \nc {\eeq} {\nonumber \end{eqnarray}}
  \nc {\eeqn}[1] {\label {#1} \end{eqnarray}}
  \nc {\eol} {\nonumber \\}
  \nc {\eoln}[1] {\label {#1} \\}
  \nc {\ve} [1] {\mbox{\boldmath $#1$}}
  \nc {\mrm} [1] {\mathrm{#1}}
  \nc {\half} {\mbox{$\frac{1}{2}$}}
  \nc {\thal} {\mbox{$\frac{3}{2}$}}
  \nc {\la} {\mbox{$\langle$}}
  \nc {\ra} {\mbox{$\rangle$}}
  \nc {\etal} {\emph{et al.\ }}
  \nc {\eq} [1] {(\ref{#1})}
  \nc {\Eq} [1] {Eq.~(\ref{#1})}
  \nc {\Ref} [1] {Ref.~\cite{#1}}
  \nc {\Refc} [2] {Refs.~\cite[#1]{#2}}
  \nc {\Sec} [1] {Sec.~\ref{#1}}
  \nc {\chap} [1] {Chapter~\ref{#1}}
  \nc {\anx} [1] {Appendix~\ref{#1}}
  \nc {\tbl} [1] {Table~\ref{#1}}
  \nc {\fig} [1] {Fig.~\ref{#1}}
  \nc {\ex} [1] {$^{#1}$}
  \nc {\Sch} {Schr\"odinger }
  \nc {\flim} [2] {\mathop{\longrightarrow}\limits_{{#1}\rightarrow{#2}}}
  \nc {\textdegr}{$^{\circ}$}
\title{Analysis of Coulomb breakup experiments of \ex{8}B with
a dynamical eikonal approximation}
\author{G.~Goldstein}
\email{gerald.goldstein@ulb.ac.be}
\author{P.~Capel}
\email{pierre.capel@centraliens.net}
\author{D.~Baye}
\email{dbaye@ulb.ac.be}
\affiliation{Physique Quantique, C.P. 165/82 and 
Physique Nucl\'eaire Th\'eorique et Physique Math\'ematique, C.P. 229,
Universit\'e Libre de Bruxelles, B 1050 Brussels, Belgium}
\date{\today}
\begin{abstract}
Various measurements of the Coulomb breakup of \ex{8}B are
analysed within the dynamical eikonal approximation
using a single description of \ex{8}B.
We obtain a good agreement with experiment for different observables measured
between 40 and 80~MeV per nucleon.
A simple \ex{7}Be-p potential model description of \ex{8}B seems sufficient 
to describe all observables.
In  particular, the asymmetry in parallel-momentum distributions
due to $E1$-$E2$ interferences is well reproduced without any scaling.
The projectile-target nuclear interactions
seem negligible if data are selected at forward angles.
On the contrary, like in previous analyses we observe a significant influence
of higher-order effects.
The accuracy of astrophysical $S$ factors 
for the $^7$Be$(p,\gamma$)$^8$B reaction at stellar energies 
extracted from breakup measurements therefore seems difficult to evaluate.
\end{abstract}
\maketitle
\section{Introduction}
The $^7$Be$(p,\gamma$)$^8$B radiative capture reaction is one of the key 
reactions for understanding neutrino properties \cite{Bah89}. 
Indeed, the main part of high energy neutrinos emitted by the sun arise 
from this reaction. 
Its properties determine our knowledge of neutrino oscillations. 

The capture of protons by $^7$Be has been studied by many direct measurements 
(see Refs.~\cite{Ham01,Bab03l,Jun03} and references therein). 
However the difficulty of these measurements and the scatter of their results 
has raised interest in indirect methods where the time-reversed reaction 
is simulated by virtual photons in the Coulomb field of a heavy nucleus 
\cite{BBR86}. 
The radiative capture cross section can be extracted if one assumes that 
the breakup is due to an E1 virtual photon and occurs in a single step. 
Several experiments have studied the breakup of $^8$B at different energies 
\cite{Mot94,Kik97,Dav98,Gui00,Dav01l,Sch06}.

Though appealing, the breakup method also faces a number of difficulties. 
First, while the reaction $^7$Be$(p,\gamma$)$^8$B is dominated 
by an E1 transition, the E2 contribution to the breakup of $^8$B 
is not negligible \cite{EB96}. 
Second, higher-order effects, i.e.\ transitions from the initial bound 
state into the continuum through several steps are not negligible \cite{EB96}. 
Finally, the nuclear interactions between $^8$B and the target may interfere 
with the Coulomb interaction.
Therefore elaborate reaction theories must be used to interpret 
the experimental data. 
Such calculations have been performed with perturbation theories 
\cite{TB94,EB96}, DWBA \cite{ST99},
semi-classical methods \cite{EB96,TWB97,DT03,EBS05}, 
and the continuum-discretized coupled channels (CDCC) method 
\cite{TNT01,MTT02,Oga06}. 

A few years ago, we have developed a technique of resolution of the 
semi-classical time-dependent Schr\"odinger equation on a mesh 
in spherical coordinates \cite{MB99,MB01,CBM03c}. 
This calculation on a mesh avoids partial wave expansions of the 
wave functions and a multipole expansion of the interaction. 
We have used this method to explore certain aspects of the extraction 
of the astrophysical $S$ factor from breakup cross sections 
\cite{MB01,CB05}. 

Recently, we have developed a purely quantal method based on the 
same semi-classical code, the dynamical eikonal approximation (DEA)
\cite{BCG05,GBC06}. 
This method allows taking into account purely quantal effects 
such as interferences, as well as calculating differential cross 
sections. 

The purpose of this paper is to analyze $^8$B breakup on a lead target 
within the DEA.
Until now, theoretical works have focused on a single experiment, 
i.e.\ the RIKEN experiment at 52 MeV/nucleon \cite{Kik97} 
or the MSU experiments at 44 and 81 MeV/nucleon \cite{Dav98},
and 83~MeV/nucleon \cite{Dav01l}. 
Here we address both experiments within exactly the same model, 
without any fit of parameters. 
The theoretical model will then serve as a link between those experiments. 
We do not consider the GSI experiment \cite{Sch06} because it is performed 
at a much higher energy where relativistic effects become important. 
The present study is nonrelativistic, except for kinematical effects, 
before and after the reaction, which are treated relativistically. 
Neither will we analyse the Notre-Dame experiment \cite{Gui00}.
It has been performed at sub-Coulomb energies, where
the DEA is not reliable.

In \Sec{DE}, we summarize the DEA, 
describe the cross section formulas in the center-of-mass
and laboratory frames 
and present our treatment of relativistic corrections. 
The condition of the calculations, including the description of
the projectile and numerical inputs, are given in \Sec{conditions}.
In \Sec{calc}, the results are presented and commented, and the accuracy 
of the extraction of the astrophysical $S$ factor is discussed. 
\Sec{conclusion} is devoted to concluding remarks.  
\section{Dynamical eikonal approximation}\label{DE}
\subsection{Principle}\label{principle}
We are interested in describing a reaction in which a two-body projectile $P$ made up of 
a structureless core $c$, with mass $m_c$ and charge $Z_c e$, and a structureless fragment $f$, 
with mass  $m_f$ and charge $Z_f e$, is broken up after its interaction with 
a target $T$, with mass $m_T$ and charge $Z_T e$. 
Since the target state remains unchanged, this process is also called elastic breakup. 
We work in Jacobi coordinates where $\ve{R}$ is the coordinate of the center of mass 
of the projectile with respect to the target 
and $\ve{r}$ is the coordinate of the fragment with respect to the core. 
The corresponding momenta are $\ve{P}$ and $\ve{p}$, respectively. 
The projectile is described by an internal Hamiltonian 
\beq
H_0 = \frac{p^2}{2\mu_{cf}} + V_{cf} (\ve{r}),
\eeqn{intH}
where $\mu_{cf}$ is the core-fragment reduced mass. 
Hamiltonian $H_0$ is composed of the kinetic energy operator for the relative motion 
between core and fragment and of the core-fragment interaction potential. 
The potential $V_{cf}$ contains an angular-momentum dependent central term 
(including a Coulomb interaction) and a spin-orbit term involving the fragment spin. 
The spin of the core is neglected. 
The eigenstates of $H_0$ with energy $E$ are denoted
as $\phi_{ljm}(E,\ve{r})$, 
where $j$ is the angular momentum resulting from the coupling of the orbital momentum $l$ 
with the fragment spin $I$ and $m$ is its projection. 
The projectile is initially in its ground state $\phi_{l_0j_0m_0}(E_0,\ve{r})$ 
and the asymptotic projectile-target relative velocity is given by $v$. 

With these assumptions, the system is described by the three-body Schr\"odinger equation 
\beq
\left[ \frac{P^2}{2\mu}  + H_0 + V_{cT}(\ve{r},\ve{R}) + V_{fT}(\ve{r},\ve{R}) \right] 
\Psi(\ve{r},\ve{R}) = E_T \Psi(\ve{r},\ve{R}),
\eeqn{TBSE}
where $\mu$ is the projectile-target reduced mass 
and $E_T$ is the total internal energy of the three-body system. 
Optical potentials $V_{cT}$ and $V_{fT}$ simulate the core-target 
and fragment-target interactions, respectively. 
In the DEA \cite{BCG05,GBC06}, 
after posing $\Psi(\ve{R},\ve{r}) = \exp(iKZ) \hat{\Psi}(\ve{R},\ve{r})$, 
we apply the eikonal approximation, i.e. we neglect second derivatives of 
$\hat{\Psi}$ which are small at high velocities. 
We do however not perform the adiabatic approximation, 
i.e. $H_0$ is not replaced by $E_0$. 
The resulting equation looks like the semiclassical 
time-dependent Schr\"odinger equation with straight lines trajectories 
\beq
	i \hbar \frac{\partial}{\partial t} \hat{\Psi}(\ve{r},\ve{b},t)
	= [ H_0(\ve{r}) + V_{cT}(\ve{r},\ve{b},t) + V_{fT}(\ve{r},\ve{b},t) -E_0] \hat{\Psi} (\ve{r},\ve{b},t).
\eeqn{TDSE}
In this equation, the variable $t$ is linked to the part of the quantal coordinate $\ve{R}$ parallel to 
the incident direction $Z=vt$, while the vector $\ve{b}=(b,\phi)$ represents the transverse part of $\ve{R}$. 
Its norm $b$ can be assimilated to the semiclassical impact parameter. 
In theory, this equation must be solved for all values of $b$ and $\phi$, 
but we will see that only one arbitrary value of $\phi$ is actually needed. 

Let $\hat{\Psi}^{(m_0)}(\ve{r},b,t)$ be a particular solution of equation (\ref{TDSE}) corresponding to the initial condition 
$\hat{\Psi}^{(m_0)}(\ve{r},b,t) \mathop{\rightarrow} \limits_{t \rightarrow -\infty} \phi_{l_0j_0m_0}(E_0,\ve{r})$ 
and to the particular orientation $\phi=0$. 
In the basis $|lIjm\ra$ coupling the orbital momentum $l$ and the fragment spin $I$, 
the asymptotic form of this solution reads 
\beq
\lim_{t \rightarrow +\infty} \hat{\Psi}^{(m_0)}(\ve{r},b,t) 
= \frac{1}{r} \sum_{ljm} \psi^{(m_0)}_{ljm}(r,b,0) \la \Omega_r |lIjm\ra.
\eeqn{24}
>From this solution, one can derive the solutions for $\phi \neq 0$ using
\beq
\psi^{(m_0)}_{ljm}(r,b,\phi) = e^{i(m_0-m)\phi} \psi^{(m_0)}_{ljm}(r,b,0).
\eeqn{Skljm2}

The whole information needed to extract cross sections is contained in the breakup amplitude given by \cite{GBC06}
\beq
S^{(m_0)}_{kljm}(b) = e^{i(\sigma_{l}+\delta_{lj}-l\pi/2)} \int_0^\infty u_{klj}(r) \psi^{(m_0)}_{ljm}(r,b,0) dr,
\eeqn{Skljm}
where $\sigma_{l}$ and $\delta_{lj}$ are the Coulomb and nuclear phase shifts
at positive energy $E=\hbar^2 k^2 / 2\mu_{cf}$.
The functions $u_{klj}(r)$ are the radial parts of eigenstates
of $H_0$ at energy $E$. They are normalized according to
\beq
u_{klj}(r)\flim{r}{\infty}\cos\delta_{lj}F_l(kr)+\sin\delta_{lj}G_l(kr),
\eeqn{norm}
where $F_l$ and $G_l$ are the Coulomb functions \cite{AS70}.

\subsection{Cross sections in the center of mass frame}
In the center of mass frame of the projectile and the target, the breakup transition matrix element 
for final projectile-target wave vector $\ve{K'}=(K',\Omega)=(K',\theta,\varphi)$ 
within the DEA is given by \cite{GBC06,B8bdp}
\beq
	T_{fi}(\ve{k},\Omega) &=& 
	i 8\pi^2 \frac{\hbar v}{k}
	\sum_{ljm} (lIm-\nu\nu|jm) Y_l^{m-\nu}(\theta_k,\varphi_k) \nonumber \\
	& & i^{-|m-m_0|} e^{i(m_0-m)\varphi} 
	\int_0^\infty b db J_{|m-m_0|}(qb) S^{(m_0)}_{kljm}(b),
\eeqn{TDE}
where $\nu$ is the projection of the fragment spin of the final state, 
$(\theta_k,\varphi_k)$ is the emission direction of the fragments of the 
projectile, and $q\approx2K\sin(\theta/2)$ is the transfered momentum. 
The phase in equation (\ref{Skljm2}) ensures the rotational symmetry around the $Z$-axis 
since the modulus of $T_{fi}$ depends on $\varphi-\varphi_k$. 

After integration over $\theta_k$, and $\varphi_k$,
these transition matrix elements lead to the differential
cross section given by \cite{GBC06}
\beq
\frac{d\sigma}{dE d\Omega} 
= K K' \frac{2\mu_{cf}}{\pi \hbar^2 k} \frac{1}{2j_0+1} \sum_{m_0} \sum_{ljm} 
\left| \int_0^\infty b db J_{|m-m_0|}(qb) S^{(m_0)}_{kljm}(b) \right|^2,
\eeqn{30a}
where $K=\mu v / \hbar$.
The sums over the $2j_0+1$ values of $m_0$ appearing in formulas (\ref{30a}) and (\ref{CS1})
can be reduced to only positive or negative values of $m_0$ by using the following symmetry property
\beq
	S^{(-m_0)}_{kljm}(b) = (-)^{l_0+j_0+m_0+l+j-m} S^{(m_0)}_{klj-m}(b).
\eeqn{sym}
\subsection{Cross sections in the laboratory frame}
Most experimental results are presented in the laboratory frame,
so that a frame transformation from the center of mass frame must be performed. 
Let $E_c$, $\Omega_c$, and $\Omega_f$ be the core energy, the core and fragment 
directions of emission, respectively. 
The cross section then reads 
\beq
	\frac{d\sigma}{dE_c d\Omega_c d\Omega_f} &=& \frac{2\pi}{\hbar v} \frac{1}{2j_0+1} \sum_{m_0\nu} |T_{fi}(\ve{k},\Omega)|^2
	\rho (E_c,\Omega_c,\Omega_f),
\eeqn{CS1}
where $\rho (E_c,\Omega_c,\Omega_f)$ is the three-body phase space factor
\cite{Fuc82} given by Eq.~(\ref{B7}) (see also \Ref{TNT01}). 
The transition matrix element $T_{fi}$ is evaluated at values of $\ve{k}$ and $\Omega$ 
linked to the values of the core momentum $\ve{p}_c$ and the fragment momentum $\ve{p}_f$ 
in the laboratory frame, by 
\beq
	\hbar \ve{k} &=& \frac{m_c}{m_P} \ve{p}_f - \frac{m_f}{m_P} \ve{p}_c \label{CV1} \\
	\hbar \ve{K'} &=& \ve{p}_c + \ve{p}_f - \frac{m_P}{m_T+m_P} \ve{p}_{tot},
\eeqn{CV2}
where $\ve{p}_{tot}=m_P \ve{v}$ is the total momentum. 
Detailed expressions of the frame transformation are given in Appendix \ref{appendixB}. 
Notice that the transition matrix element $T_{fi}$ must be calculated for each $(\ve{k},\Omega)$. 
In particular, for each couple $(k,q)$, the breakup amplitude $S^{(m_0)}_{kljm}(b)$ 
and its integrals over $b$ in Eq.~(\ref{TDE}) are recalculated. 

We are also interested by the computation of the core parallel momentum distribution, 
so that the cross section (\ref{CS1}) must be integrated three times using the following expression 
\beq
	\frac{d\sigma}{dp_{c\parallel}} =  \frac{2\pi}{m_c} \int_{|p_{c\parallel}|}^{p_{c}^{\max}} dp_{c}
	\int_0^\pi d\theta_f \sin\theta_f \int_0^{2\pi} d\Delta\varphi\ \frac{d\sigma}{dE_c d\Omega_c d\Omega_f},
\eeqn{CS2}
where $p_{c}^{\max} = |p_{c\parallel}| / \cos\theta_{c}^{\max}$ and $\Delta\varphi=\varphi_c-\varphi_f$. 
Notice that the two integrations over $\varphi_c$ and $\varphi_f$ reduce to a single integration over $\Delta\varphi$ 
since $d\sigma / dE_c d\Omega_c d\Omega_f$ is a periodic function of $\Delta\varphi$ 
[see equations (\ref{TDE}) and (\ref{B1}) to (\ref{B7})]. 
As for the cross sections in the c.m. frame, the sum over $m_0$ can also be reduced using property (\ref{sym}). 
This is the case only when the cross sections is integrated over $\Delta\varphi$ like in expression (\ref{CS2}).
\subsection{Kinematical relativistic corrections}\label{rel}
As mentioned in the Introduction, the present model is nonrelativistic like 
its basic equation (\ref{TBSE}). 
However, kinematical relativistic corrections are used before and after the reaction. 

The velocity of the projectile $v$ used for solving the time-dependent 
Schr\"odinger equation (\ref{TDSE}) is calculated using the relativistic formula 
\beq
	\frac{v}{c} = \sqrt{1-\frac{1}{(1+\frac{T_i}{m_Pc^2})^2}},
\eeqn{rel1}
where $T_i$ is the initial kinetic energy of the projectile and $m_P$ its mass.
This velocity is also used for the frame transformation. 
This is common practice. 

In order to be consistent with our velocity choice, 
we use at the end of the calculation the inverse relation
for the parallel momentum,
i.e. we associate to each observed momentum $\ve{p}_c$
a non-relativistic value $\ve{p}_c^{NR}$.
The parallel component of this nonrelativistic momentum
$p_{c\parallel}^{NR}$ at which the cross section must be computed to
simulate the actual one $p_{c\parallel}$ is
\beq
p_{c\parallel}^{NR}=
\left(\frac{1}{(m_Pc)^2}+\frac{1}{p_{c\parallel}^2}\right)^{-1/2},
\eeqn{rel2}
while its transverse component remains unchanged. 
Hence the non relativistic emission angles $\theta_c^{NR}$
giving the direction of the final momentum are
obtained according to
%transformed into 
\beq
\tan\theta^{NR}_c = \frac{p_{c\parallel}}{p^{NR}_{c\parallel}} \tan\theta_c.
\eeqn{rel3}
This reduction will in particular affect the cutoff angles. 

\section{Condition of calculations}\label{conditions}
\subsection{Description of \ex{8}B}\label{B8}
The spectrum of \ex{8}B contains only one bound state with
$J^\pi=2^+$. It is bound by a mere 137~keV in regards to
the one-proton separation.
Therefore \ex{8}B is usually seen as a valence proton loosely bound to a
\ex{7}Be core. In the dominant configuration this proton
is in a $0p3/2$ orbit coupled to the $\thal^-$ ground state
of \ex{7}Be \cite{BDT94}.
For computational reasons, we restrict ourselves here to a
simple model of \ex{8}B, in which the spin and internal structure
of the core are neglected. This description corresponds to the
Hamiltonian $H_0$ given in \Eq{intH}.
The \ex{7}Be-p potential $V_{cf}$ contains both Coulomb and
nuclear terms. The former is a point-sphere Coulomb potential
of radius $R_C=2.391$~fm. The latter is a central Woods-Saxon potential
plus a spin-orbit coupling term. Their radius and diffuseness are
$R_0=2.391$~fm and $a=0.52$~fm, respectively.
The strength of the spin-orbit coupling
term is $V_{LS}=19.59$~MeV~fm\ex{2}.
The depth of the central nuclear potential $V_0$ is adjusted in order
to reproduce the $0p3/2$ bound state at $-137$~keV.
With $\hbar^2/2\mu_{cf}=23.698$~MeV~fm\ex{2}, we obtain $V_0=44.65$~MeV.
This \ex{8}B description corresponds to a simplified version
of the model of Esbensen and Bertsch \cite{EB96}.
It has been used in previous dynamical calculations of \ex{8}B breakup
\cite{EB96,Dav01c,TNT01,MTT02}.

The simplicity of this structure model has several drawbacks.
First, since the spin of the core is neglected, it is not possible
to adjust the \ex{7}Be-p interaction to reproduce the
scattering length in both $S=1$ and $S=2$ total-spin channels.
Choosing the same $V_{cf}$ potential in all partial waves
gives $a_0=5.9$~fm, in between the (large) positive value
of the $S=1$ channel ($25\pm9$~fm) and the negative value
of the $S=2$ channel ($-7\pm3$~fm) \cite{Ang03}.
Second, the absence of core spin means that all $0^+$, $1^+$, and $3^+$ states
corresponding to the coupling of this spin with the
$3/2$ angular momentum of the proton are degenerate with the
$2^+$ ground state. In particular, the $1^+$ resonance located 632~keV
above the proton separation threshold is not reproduced by this model.
Finally, the structure of the core being neglected, no process
leading to its excitation can be simulated.

\subsection{Projectile-target interactions}\label{PT}
The nuclear interactions between the projectile constituents and
the target are simulated with optical
potentials chosen in the literature.
Unfortunately, no scattering data are available for the elastic scattering
of \ex{7}Be on Pb at energies of interest.
Following Mortimer, Thompson, and Tostevin \cite{MTT02},
we consider the potential developed by Cook \cite{Coo82}.
It is a global parametrization determined from
elastic scattering data of \ex{7}Li on various targets at
different energies.
To describe the p-Pb interaction, we use the
parametrization of Becchetti and Greenlees \cite{BG69},
neglecting the spin-orbit coupling term.

\subsection{Numerical inputs}\label{num}
The time-dependent \Sch equation obtained from the
DEA \eq{TDSE} is solved
using the numerical technique detailed in \Ref{CBM03c}.
In this technique, the projectile internal wave function
is expanded upon a three-dimensional spherical mesh.
The angular grid contains up to $N_\theta=10$ points along
the colatitude $\theta$, and $N_\varphi=19$ points along the
azimuthal angle $\varphi$. This corresponds to an angular basis
that includes all possible spherical harmonics up to $l=9$.
The radial variable $r$ is discretized on a quasiuniform mesh that
contains $N_r=800$ points and extends up to $r_{N_r}=800$~fm.
The evolution calculation is performed with a second-order
approximation of the evolution operator.
It is started at $t_{\rm in}=-20~\hbar/$MeV with the projectile
in its initial bound state, and is stopped at $t_{\rm out}=20~\hbar/$MeV
($t=0$ corresponds to the time of closest approach).
The time step is set to $\Delta t=0.04\hbar/$MeV.

The evolution calculations are performed for different values
of the transverse component $b$ of the projectile-target coordinate
(see \Sec{principle}). These values range from 0 up to 200~fm with a
step $\Delta b$ varying from 0.5~fm to 5.0~fm, depending on $b$.
The integrals over this impact parameter
appearing in Eqs.~\eq{TDE}, and \eq{30a}
are performed numerically.
To take into account the rapid variation of the Bessel function,
the values of the breakup amplitude $S_{kljm}^{(m_0)}$ are
interpolated on a uniform grid in $b$ with a step of 0.05~fm.
The convergence of these integrals is ensured by
extrapolating the $S_{kljm}^{(m_0)}$ up to $b=500$~fm
as explained in Sec.~III~A of \Ref{GBC06}.

\section{Analysis of experiments}\label{calc}
\subsection{Longitudinal momentum distribution}\label{dp}
The longitudinal momentum distribution \eq{CS2}
of \ex{7}Be obtained from the
dissociation of \ex{8}B on lead at incident energies of 44 and
81~MeV/nucleon has been measured at MSU \cite{Dav98,Dav01c}.
These measurements have been performed with the aim of evaluating
the contribution of the $E2$ strength in Coulomb breakup.
This observable is indeed the best suited
to measure the influence of the electric quadrupole transition since,
as shown by Esbensen and Bertsch \cite{EB96}, $E1$-$E2$ interferences
produce a significant asymmetry in that distribution.

\begin{figure}
\includegraphics[width=10cm]{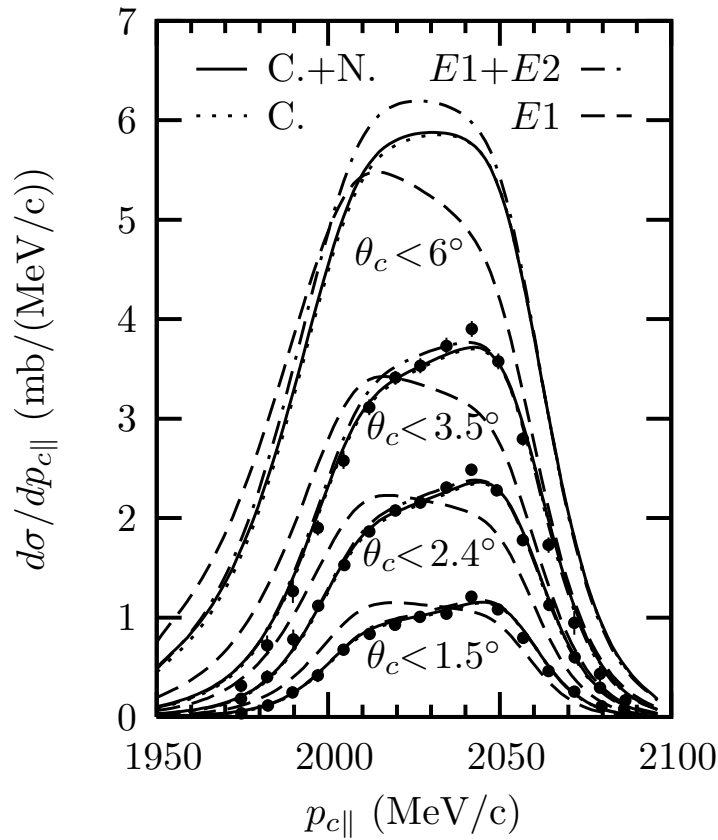}
\caption{Longitudinal momentum distributions of \ex{7}Be obtained by
dissociation of \ex{8}B on Pb at 44~MeV/nucleon.
The upper set of curves corresponds to \ex{7}Be scattering-angle cut
$\theta_c<6$\textdegr. The lower three, along with the experimental data
\cite{Dav98}, correspond to
$\theta_c<3.5$\textdegr, $2.4$\textdegr, and $1.5$\textdegr.
DEA calculations are performed using
Coulomb plus nuclear (full lines), purely Coulomb (dotted lines), 
$E1+E2$ (dash-dotted lines), and $E1$ (dashed lines) $P$-$T$ interactions.
All distributions are convoluted with the experimental
resolution of 5~MeV/$c$. For comparison with the data they are also
shifted by $-5$~MeV/$c$.}\label{f44}
\end{figure}

The results of our calculations are displayed alongside
the experimental data in \fig{f44} for the breakup of \ex{8}B
on lead at 44~MeV/nucleon. Each set of curves corresponds to
a given cut in the scattering angle $\theta_c$ of \ex{7}Be:
the upper one is obtained for $\theta_c<6$\textdegr;
the lower ones correspond to the experimental angle cuts
$\theta_c<3.5$\textdegr, $2.4$\textdegr, and $1.5$\textdegr.
To allow the comparison with experiment, our calculations
have been convoluted with the experimental resolution
of 5~MeV/$c$ \cite{Dav98}.
They have also been shifted by $-5$~MeV/$c$ so as to be centered
on the data.
It can be seen as a fine adjustment to the data
not taken into account by the relativistic corrections (see \Sec{rel}).
This shift is very small compared to the central value of the longitudinal
momentum (2025~MeV/$c$). Moreover it is equal to the experimental resolution.
The relativistic corrections \eq{rel2}-\eq{rel3} used to calculate the parallel
momentum in the laboratory frame thus seem efficient.
Finally, note that no other parameter fitting has been performed.
In particular, the magnitude of the curves is exactly
that obtained from our calculations.

The full lines (labeled C.+N.) in \fig{f44} correspond
to calculations performed
with realistic projectile-target ($P$-$T$) interactions,
i.e. using optical potentials
that simulate both Coulomb and nuclear interactions (see \Sec{PT}).
These results are in very good agreement with the data
at the three forward angle cuts.
They match both the magnitude and width of the experimental distributions.
Moreover, they reproduce fairly well the asymmetry observed by
Davids \etal \cite{Dav98}.
This is particularly true at the smallest
angle cut, where our calculations are nearly superimposed on the data.
Note that with a cutoff on $\theta_c$ at 6\textdegr,
the distribution exhibits very little asymmetry.

These results confirm the validity of the DEA
to describe the dissociation of loosely-bound projectiles at intermediate
energies, as already observed for the dissociation
of \ex{11}Be \cite{BCG05,GBC06}.
Moreover they imply that the simple model of \ex{8}B is
sufficient to describe this breakup observable.

To better understand the reaction mechanism, as well as analyse
the role of each interaction in the dissociation, the calculations
are performed with different choices of $P$-$T$ potentials.
First, we consider a purely Coulomb interaction.
These results are shown as dotted lines in \fig{f44} (labeled C.).
They are barely distinguishable from the previous ones
(even for $\theta_c<6$\textdegr),
which indicates that $P$-$T$ nuclear interactions
are negligible when data are restricted to forward angles.
The calculations can therefore be performed with a purely
Coulomb interaction as done by Davids and Typel \cite{DT03}.
This avoids the uncertainty due to the choice of
optical potentials.

The insensitivity to optical potentials also indicates
that there is no absorption from the elastic-breakup channel
at forward angles.
This justifies the use of reaction models
in which only elastic breakup is considered to analyse
the inclusive measurements of Davids \etal \cite{Dav98},
where the valence proton is not detected in coincidence
with the \ex{7}Be core.

To evaluate the influence of the different multipoles of
the Coulomb interaction, we also perform DEA calculations
considering only the dipole term ($E1$), and the sum
of the dipole and quadrupole terms ($E1+E2$).
The results obtained with the purely $E1$ interaction
(dashed lines) confirm the previous analyses
\cite{EB96,Dav98,Dav01c,MTT02,DT03}: $E1$ strength alone is not able to
reproduce the data. In particular, this calculation
exhibits a reverse asymmetry compared with the data,
even at very forward angles.
The quadrupole term of the Coulomb interaction
is therefore mandatory to correctly interpret the data.
We see indeed that the results obtained with
both dipole and quadrupole terms (dash-dotted lines)
are in better agreement with the data.
Note that we still observe some slight difference with the
full calculation. This difference becomes negligible at very
small $\theta_c$. It is probably due to higher multipoles.
In a semiclassical viewpoint, their contributions
indeed decrease at small scattering angle
(i.e. large impact parameter) \cite{AW75}.

\begin{figure}
\includegraphics[width=10cm]{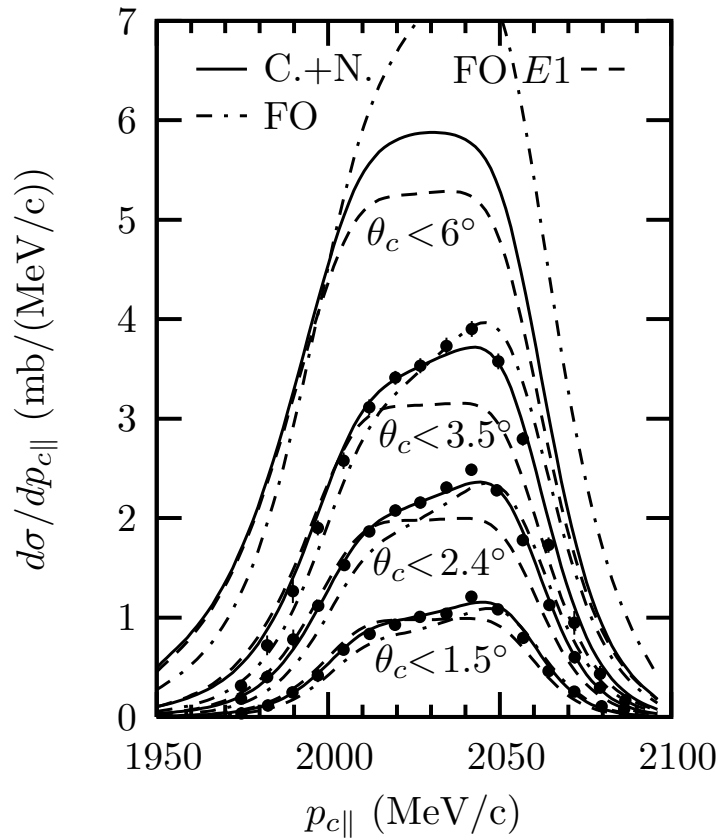}
\caption{Longitudinal momentum distributions of \ex{7}Be obtained by
dissociation of \ex{8}B on Pb at 44~MeV/nucleon.
Comparison of DEA and first-order (FO) calculations
performed for angular cuts at 6\textdegr, 3.5\textdegr,
2.4\textdegr, and 1.5\textdegr.
Experimental data are from \Ref{Dav98}.
DEA calculations are performed using
Coulomb plus nuclear (full lines).
First-order calculations are performed with $E1+E2$ (dash-dotted lines)
and $E1$ (dashed line) strengths.
All distributions are convoluted with the experimental
resolution of 5~MeV/$c$, and shifted by $-5$~MeV/$c$.}\label{f44FO}
\end{figure}

We now compare the DEA to the
first-order perturbation theory (FO) \cite{AW75}.
In the latter approximation, the time-dependent \Sch equation \eq{TDSE}
is solved perturbatively assuming straight-line trajectories,
and purely Coulomb $P$-$T$ interaction.
This potential is expanded in a series of multipoles $E\lambda$,
whose contributions to the breakup amplitude
$S^{(m_0)E\lambda}_{kljm}(b)$ is given
within the far-field approximation by \Eq{SkljmE1E2}.
The corresponding parallel-momentum distributions
are computed assuming the classical relation between the
impact parameter $b$ and the scattering angle $\theta$
of the \ex{8}B center of mass \cite{AW75}.
The comparison is made in \fig{f44FO}.
As in \fig{f44}, the full lines depict the DEA results
obtained considering both Coulomb and nuclear interactions
between the projectile and the target (C.+N.).
The dashed lines correspond to the first-order calculations
performed with only the dipole term of the Coulomb interaction (FO $E1$).
The first-order calculations obtained using dipole and quadrupole
strengths are plotted as dot-dashed lines (FO).

At the first-order approximation, the $E1$ distribution shows no asymmetry.
This is very different from the dynamical calculation, which shows
a reverse asymmetry compared with the data (see dashed curves in \fig{f44}).
The difference is due to higher-order effects,
which are neglected at the first-order of the perturbation theory.
Such a substantial effect of higher-orders
is in complete agreement with previous analyses that indicate
a strong coupling between partial waves inside the continuum \cite{CB05,EBS05}.
An analysis of this experiment ignoring these effects
seems therefore unrealistic.

Once the quadrupole term is included, even first-order theory
leads to an asymmetric distribution. However, as already noted
by previous studies, this is too strong an asymmetry
\cite{EB96,Dav01c,MTT02,DT03}: higher-order
effects tend to reduce the asymmetry.
This tendency is of course to be related to the reversal
of the asymmetry observed for the purely $E1$ case.
It confirms the necessity
of taking into account dynamical effects to analyse the data.

\begin{figure}
\includegraphics[width=10cm]{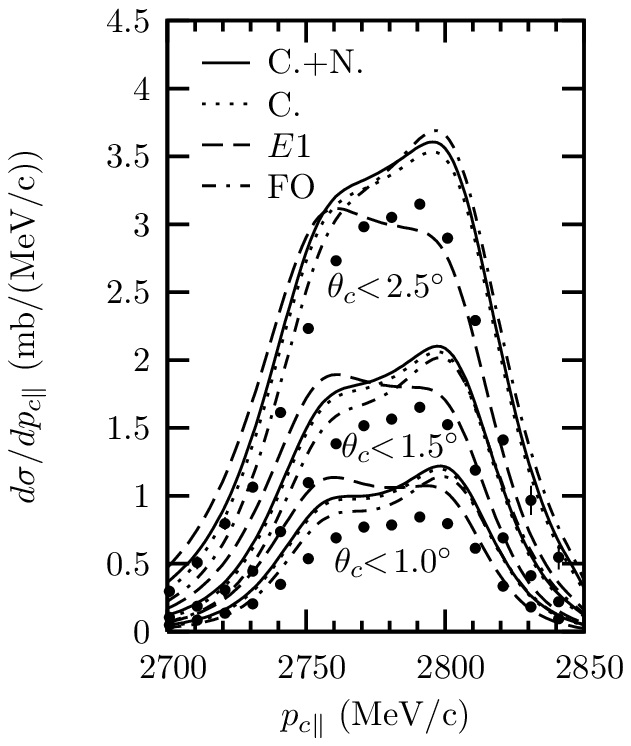}
\caption{Longitudinal momentum distributions of \ex{7}Be obtained by
dissociation of \ex{8}B on Pb at 81~MeV/nucleon.
The sets of results correspond to \ex{7}Be scattering-angle cuts
at 2.5\textdegr, 1.5\textdegr, and 1.0\textdegr.
Experimental data are from \Ref{Dav98}.
DEA calculations are performed using
Coulomb plus nuclear (full lines), purely Coulomb (dotted lines), and
$E1$ (dashed lines) $P$-$T$ interactions.
First-order calculations are performed with $E1+E2$ (dash-dotted lines)
strengths.
All distributions are convoluted with the experimental
resolution of 5~MeV/$c$, and shifted by $-9$~MeV/$c$.}\label{f81}
\end{figure}

We perform a similar analysis of the 81~MeV/nucleon data
of Davids \etal \cite{Dav98}.
\fig{f81} depicts the results of our calculations alongside
the measurements.
As for the 44~MeV/nucleon case, our longitudinal momentum distributions
have been convoluted with the experimental resolution
of 5~MeV/$c$ \cite{Dav98}.
We also need to shift our calculations to center them on the data.
This shift ($-9$~MeV/$c$) is larger than for the 44~MeV/nucleon
distributions, probably denoting larger relativistic
corrections, which are less well simulated by our
purely kinematical corrections (see \Sec{rel}).
However it remains minor when compared with the average momentum
and the experimental resolution.

The results of the full calculations (i.e. containing both Coulomb and
nuclear $P$-$T$ interactions) are plotted as full lines in \fig{f81}.
The agreement with experiment is less good than at 44~MeV/nucleon.
The magnitude of the calculations is 10 to 25~\% too high,
depending on the angle cut,
and its width is also larger than the experimental one.
This could merely denote larger
relativistic effects at this larger incident energy.
It could also result form uncertainty in the data.
The very good agreement with the 44~MeV/nucleon data
might indeed be fortuitous, while the discrepancy observed
at 81~MeV/nucleon might be more representative of the
experimental uncertainty (e.g. on the angle cut).
Nevertheless, our distributions correctly reproduce the
slope at the center of the distribution, which is
usually the main concern of previous studies \cite{Dav98,Dav01c,MTT02,DT03}.

As in the previous case, there is no significant influence
of the nuclear interactions between the projectile components
and the target. The calculations performed with
purely Coulomb $P$-$T$ interactions (dotted lines)
are indeed very similar to those including nuclear optical potentials.
In particular, all distributions have about the same width.
Therefore the nuclear interaction cannot explain the too narrow
widths obtained in \Ref{DT03}, in contradiction with the authors'
explanation.

The results obtained using only the $E1$ strength are depicted
as dashed lines. Similarly to the previous case, we observe
a reverse asymmetry when compared to the experiment.
This is again a signature of the role in the dissociation
process of both higher multipoles of the Coulomb interaction, and
dynamical effects.
This is confirmed by calculations performed at the first-order
using both dipole and quadrupole strengths
(dash-dotted lines). We indeed obtain a larger
asymmetry than the fully dynamical calculation.

The calculations performed at 44 and 81~MeV/nucleon
show both an asymmetry in the \ex{7}Be longitudinal-momentum
distribution.
In agreement with previous studies \cite{EB96,Dav01c,MTT02,DT03},
this asymmetry is found to be mainly due to interferences
between dipole and quadrupole terms of the $P$-$T$
Coulomb interaction. As already observed \cite{EB96,Dav01c,MTT02,DT03}
this asymmetry is reduced by higher-order effects.
Using the simple description of \ex{8}B of Esbensen and Bertsch \cite{EB96}
within the DEA \cite{BCG05,GBC06}, we
obtain a very good agreement with the experimental data of Davids \etal
\cite{Dav98}. In particular, the asymmetry of the distribution
is well reproduced at both energies.
Contrarily to what has been suggested in previous calculations
\cite{Dav98,Dav01c,MTT02,DT03}, no scaling of the $E2$ strength
is needed to explain the data.
This suggests that the electromagnetic strengths given by
the \ex{8}B model of Esbensen and Bertsch is sufficient
to describe the parallel-momentum distribution of the core
obtained through Coulomb dissociation.
To investigate further on the validity of this model, we
now turn to other breakup observables.

\subsection{Energy distribution}\label{dE}
Consecutively to the aforementioned experiments,
the breakup of \ex{8}B on lead has been
remeasured at MSU for an 83~MeV/nucleon incident beam \cite{Dav01l}.
Contrarily to the previous measurements, this one
is exclusive in the sense that both the \ex{7}Be core and the
valence proton are detected in coincidence.
This enabled Davids \etal to obtain the breakup cross
section as a function of the \ex{7}Be-$p$ relative energy.
The aim of this experiment was to extract the $B(E1)$ from
this cross section using the first-order perturbation
theory. This $E1$ strength was subsequently used to infer
the astrophysical factor of the $^7{\rm Be}(p,\gamma)\,^8{\rm B}$
radiative capture at solar energy $S_{17}(0)$.
Knowing the $E2$ strength not to be negligible, the authors
used the $E2$ strength extracted from their previous measurements
of the \ex{7}Be longitudinal-momentum distribution \cite{Dav98}.
In the present study, we consider these data as another breakup
observable to which to compare our calculations.

\begin{figure}
\includegraphics[width=14cm]{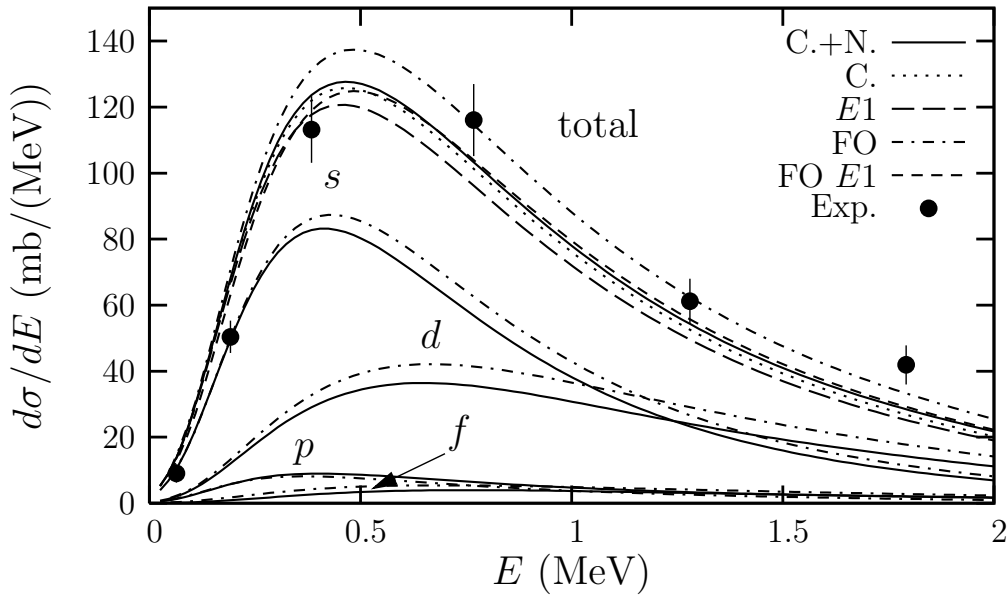}
\caption{Breakup of \ex{8}B on Pb at 83~MeV/nucleon:
relative energy distribution limited at forward \ex{8}B scattering angles
($\theta\le 1.77$\textdegr).
Dynamical calculations are performed using
Coulomb plus nuclear (full lines), purely Coulomb (dotted line), and
$E1$ (long-dashed line) $P$-$T$ interactions.
First-order calculations are performed with $E1+E2$ (dash-dotted lines)
and $E1$ (short-dashed line) strengths.
Some partial-wave contributions are also shown.
Experimental data are from \Ref{Dav01l}.
All distributions are convoluted with the experimental
resolution [see \Eq{conv}].}\label{f83}
\end{figure}

Energy distributions obtained from the DEA
are displayed in \fig{f83}. Since the difference in incident energy
is small, they are obtained from the 81~MeV/nucleon calculations.
Like the experiment, they are limited to a scattering angle of the \ex{8}B
center of mass $\theta$ smaller than $1.77$\textdegr\ \cite{Dav01l}.
For comparison with the data, the theoretical distributions
($d\sigma^{\rm th}/dE$) have been
convoluted with the experimental energy resolution \cite{Dav07p}
\beq
\frac{d\sigma^{\rm conv}}{dE}(E)=\int_0^\infty
\frac{1}{\sqrt{2\pi}a(E')}
\exp\left[-\frac{(E-E')^2}{2a(E')^2}\right]
\frac{d\sigma^{\rm th}}{dE}(E') dE',
\eeqn{conv}
where the energy-dependent width $a$ reads
\beq
a(E)=\left\{\begin{array}{ll}
0.2072\sqrt{E}-0.0145\ \ \ \ \ & \mbox{if $E>64$~keV,}\\
            0.038 & \mbox{otherwise}.
\end{array}
\right.
\eeqn{aE}
In Eqs.~\eq{conv}, and \eq{aE}, the energies $E$ and $E'$,
and the width $a$, are expressed in MeV.

The full lines correspond to a calculation
performed using realistic $P$-$T$ interactions containing
both Coulomb and nuclear potentials (see \Sec{PT}).
The upper curve is the total cross section, while the
lower curves correspond to partial wave contributions.
The agreement  with the experimental data is fair if
one considers the very simple model of \ex{8}B used here.
The maximum of our theoretical distribution seems to be
located slightly too low in energy in comparison with the data.
The magnitude of our distribution, however, is similar to
the experimental one.
This suggests some contradiction between these
data and those obtained at 81~MeV/nucleon, compared to which
our calculations are too large by 10--25\% (see \fig{f81}).

As in \Sec{dp}, we analyse the sensitivity of our calculation to
the projectile-target interactions.
The cross section obtained using purely Coulomb $P$-$T$ interactions
is shown as a dotted line.
As for the longitudinal-momentum distributions, the difference
with the calculation including both Coulomb and nuclear
interactions is negligible. This confirms that
for \ex{8}B dissociation on a heavy target at intermediate energies,
nuclear interactions can be neglected when observables
are limited to forward angles.

The result obtained with only the dipole term of the Coulomb
interaction (long-dashed line) confirms the less noticeable influence
of higher multipoles in this observable.
The shape is indeed similar whether the full interaction
or only the $E1$ strength is considered.

In \fig{f83}, we also compare our dynamical calculations
with the first-order perturbation theory (dash-dotted lines).
In the latter, the $P$-$T$ interaction is purely Coulomb and is limited
to its dipole and quadrupole terms. The scattering-angle
cut is simulated by an impact-parameter cutoff at $b_{\rm min}=30$~fm,
as suggested in \Ref{Dav01l}.
We observe a reduction between the first-order
and dynamical calculations (9\% at 0.5~MeV).
A better agreement between the first-order and DEA total cross sections
can be found using an impact-parameter cutoff at $b_{\rm min}=34$~fm,
which corresponds to the angular cut at 1.77\textdegr\
through the classical relation between $b$ and $\theta$ \cite{AW75}.
In both cases significant higher-order effects are at play.
We indeed observe that the variation between first-order
and DEA is not the same in all partial waves
(see lower curves in \fig{f83}).
While both dominant $s$ and $d$ contributions are
reduced in the DEA calculation, the $p$ contribution
is slightly increased.
In agreement with our previous analysis \cite{CB05},
we interpret this as couplings between different partial
waves in the continuum.
The variations observed here suggest that $E1$-$E1$
second-order transitions
depopulate the $s$ and $d$ waves in the continuum
towards the $p$ and $f$ waves, where they interfere
with first-order $E2$ transitions.
As already seen in \Ref{CB05}, this effect varies with
the relative energy $E$.
It seems thus hazardous to
model these higher-order effects as a mere reduction of the $E2$ strength
within the first-order perturbation theory, as suggested
in previous works \cite{Dav01l,Dav01c,DT03}.
These results confirm earlier studies \cite{EBS05,CB05},
which show that higher-order effects and $E2$ transitions
interfere in \ex{8}B Coulomb breakup,
even at intermediate incident energies, and
forward scattering angles.

\begin{figure}
\includegraphics[width=14cm]{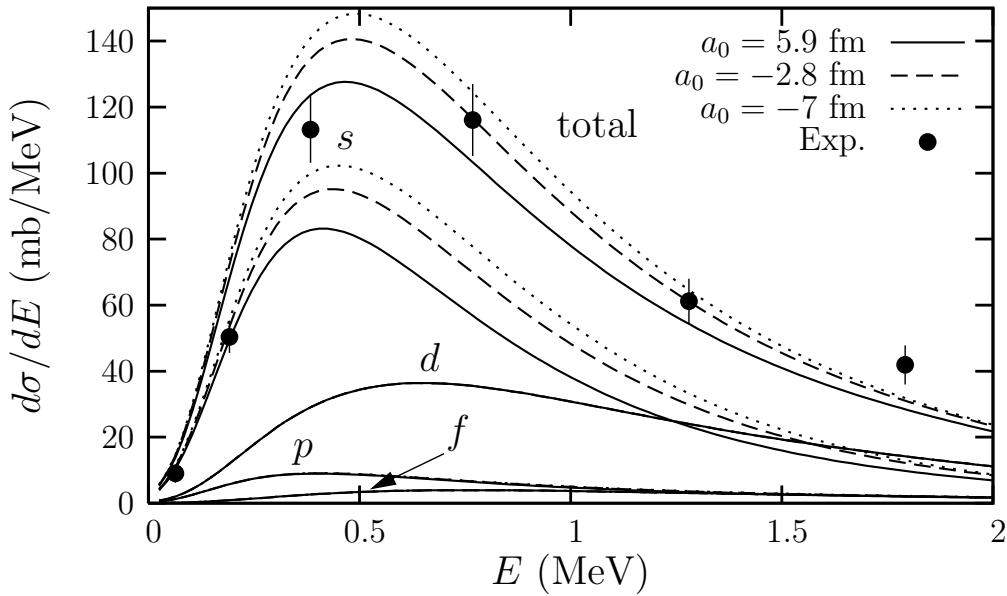}
\caption{Breakup of \ex{8}B on Pb at 83~MeV/nucleon:
influence of scattering length on breakup.
Dynamical calculations with $a_0=5.9$~fm (initial potential; full lines),
$a_0=-2.8$~fm (dashed lines), and $a_0=-7$~fm (dotted lines).
Experimental data are from \Ref{Dav01l}.
All distributions have been convoluted with the experimental
resolution [see \Eq{conv}].}\label{f83a0}
\end{figure}

Albeit fair, the agreement we obtain here between theory and
experiment is less good than for the parallel-momentum distribution.
One explanation for this difference might be the larger sensitivity
of the energy distribution to the projectile continuum.
Being integrated over the energies, the parallel-momentum
distribution \eq{CS2} might indeed be less sensitive to that part
of the projectile model.
To investigate that possibility, we perform two additional calculations
with \ex{7}Be-$p$ potentials adjusted in the $s$ wave to
reproduce various scattering lengths $a_0$.
The potential is kept unchanged in all other partial waves.
The corresponding energy distributions, as the major
partial wave contributions are displayed in \fig{f83a0}.

We first consider $a_0=-2.8$~fm (dashed lines), which corresponds
to the weighted average value between the scattering lengths for the
total-spin channels $S=1$ and $S=2$ suggested in \Ref{Bay00}.
Second we fit the potential to $a_0=-7$~fm (dotted lines), which is the
value measured for the $S=2$ channel \cite{Ang03}.
Both calculations are performed considering Coulomb plus
nuclear $P$-$T$ interactions.
The only variation we observe from the initial calculation
($a_0=5.9$~fm; full lines),
is a significant increase of the magnitude of the $s$ contribution.
The shape of that contribution is similar for all potential choices.
No significant change is observed in the other contributions.
This result confirms the influence of the description
of the projectile continuum upon breakup calculations \cite{TB05,CN06}.
However, even adjusting scattering lengths on realistic
values does not explain the difference observed between our
calculations and experimental data.
We suspect this observable to be more sensitive to
the projectile description than momentum distributions.
A more realistic model of \ex{8}B, which takes into account
the spin of the core, and reproduces the measured scattering
lengths, might explain the slight shift observed in \fig{f83}.
However, such a model is still too time consuming for the DEA.

\subsection{Angular distribution}\label{dt}
In order to complete this analysis, we use the DEA
to compute angular distributions
and compare them to the data obtained by Kikuchi \etal
at RIKEN \cite{Kik97}.
In this experiment, the cross section for the breakup \ex{8}B on lead
at 52~MeV/nucleon
has been measured as a function of the scattering angle $\theta$
of the \ex{8}B center of mass.
The main reason for this experiment was to
evaluate the contribution of $E2$ transitions
in the Coulomb breakup of \ex{8}B.

\begin{figure}
\includegraphics[width=13cm]{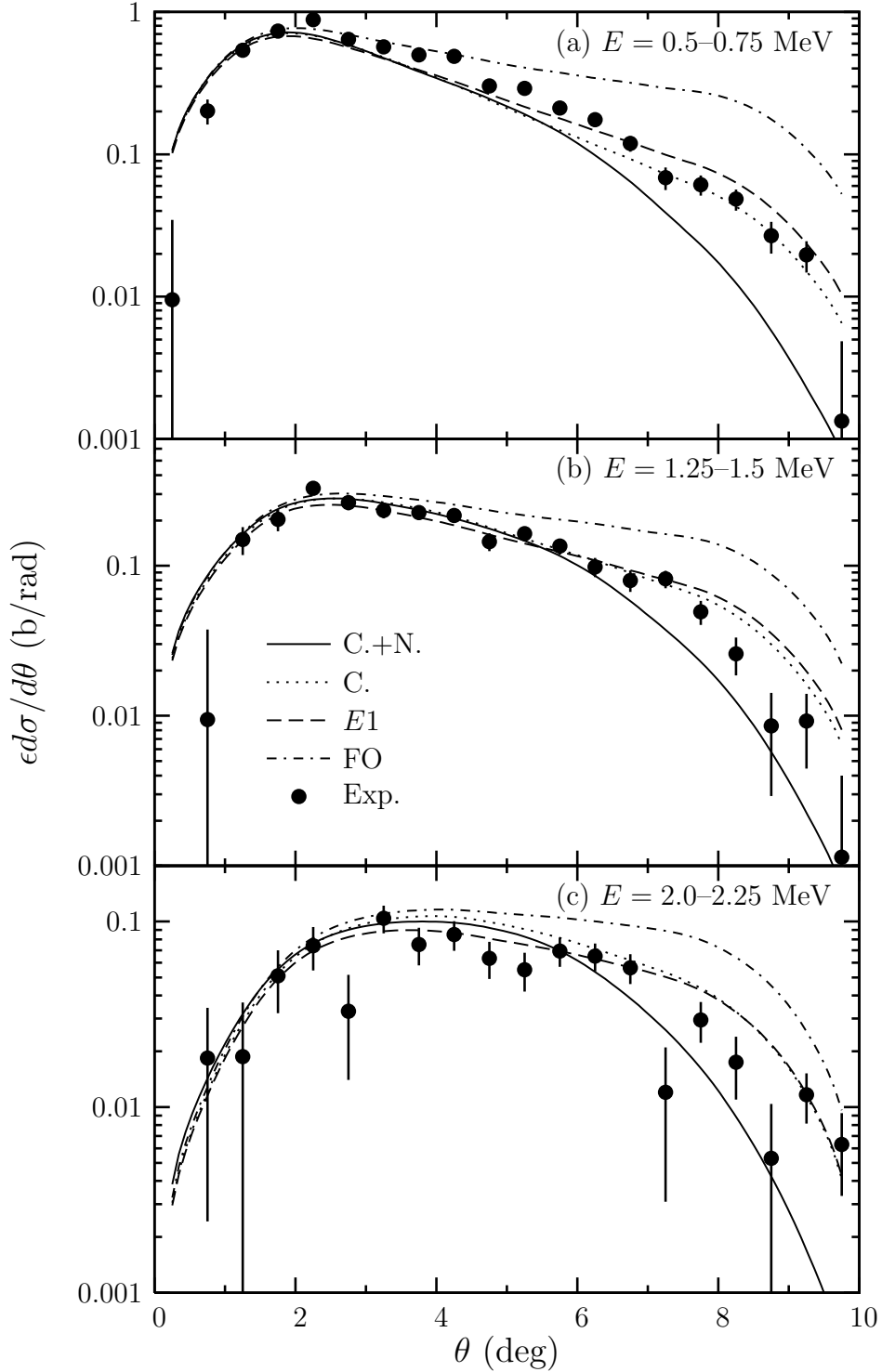}
\caption{Breakup of \ex{8}B on Pb at 52~MeV/nucleon:
angular distribution as a function of \ex{8}B center-of-mass
scattering angle $\theta$.
Three ranges for the \ex{7}Be-$p$ relative energy are considered:
(a)~$E=0.5$--0.75~MeV, (b)~$E=1.25$--1.5~MeV, and
(c)~$E=2.0$--2.25~MeV.
DEA calculations are performed using
Coulomb plus nuclear (full lines), purely Coulomb (dotted line), and
$E1$ (dashed line) $P$-$T$ interactions.
First-order calculations performed with $E1+E2$ strengths
are shown as dash-dotted lines.
Experimental data are from \Ref{Kik97}.
All distributions are convoluted with the experimental
resolution.}\label{f52}
\end{figure}

The experimental data are displayed in \fig{f52}.
The three parts correspond to different bins of
\ex{7}Be-$p$ relative energy after breakup:
(a) $E=0.5$--0.75~MeV,
(b) $E=1.25$--1.5~MeV, and
(c) $E=2.0$--2.25~MeV.
Alongside the data are shown the results of our calculations.
To allow a comparison with the data, they have been
filtered by the experimental resolution provided
by the authors of \Ref{Kik97}.

The full lines correspond to calculations performed with
$P$-$T$ interactions containing both Coulomb
and nuclear potentials.
The agreement with the data is fair, in particular at
small scattering angle (i.e. $\theta<6$\textdegr).
Indeed both the magnitude and general feature of the
data are well reproduced by our calculations.
Note that no parameter adjustment has been done to
fit the data.
At larger angle, the theoretical calculation drops
faster than the measurements.
This could be due to an inappropriate choice of optical potentials
to simulate the $P$-$T$ nuclear interactions, which indeed
are significant only at large scattering angle (see below).
Another explanation of this discrepancy could be the
filtering with the experimental resolution. Following
Ogata \etal \cite{Oga06}, it has been devised assuming the breakup
of \ex{8}B to occur only through an $s$ continuum state.
As shown by these authors this is not the case at large scattering
angle \cite{Oga06}. Therefore the quantitative comparison between
theory and experiment may not be significant for too large $\theta$.

Note that our calculations are in perfect agreement with those
of \Ref{Oga06} (see Fig.~3 of that reference) although different
reaction models and different \ex{8}B descriptions are considered.
This agreement validates both calculations.
It also suggests that the details of the \ex{8}B description
have but little effect on these calculations.
As for the parallel-momentum distributions (see \Sec{dp}),
a simple two-body model of \ex{8}B seems sufficient to reproduce
this breakup observable.

With the aim of analysing the influence of the nuclear $P$-$T$
interactions, we also perform the dynamical calculation considering
purely Coulomb $P$-$T$ potentials. The corresponding angular distributions
are displayed as dotted lines in \fig{f52}.
As observed for the parallel-momentum and energy distributions,
the difference with the Coulomb plus nuclear calculation is negligible
at forward angles (i.e. $\theta<4$--6\textdegr).
At larger angle on the contrary, the difference is more significant.
The elastic breakup is strongly reduced in the Coulomb plus nuclear case,
probably due to the absorption terms of the optical potentials. 
As mentioned earlier, this attenuation is not observed experimentally.
It could be due to an inappropriate choice of the optical potentials,
or to uncertainty in the filtering technique.
The angle at which the difference appears between the purely Coulomb
and Coulomb plus nuclear calculations decreases at larger relative
energy $E$. This is in agreement with previous calculations, where
it has been shown that nuclear potentials affect more significantly
the breakup cross section at large relative energy \cite{TS01r,CBM03c}.

\begin{figure}
\includegraphics[width=14cm]{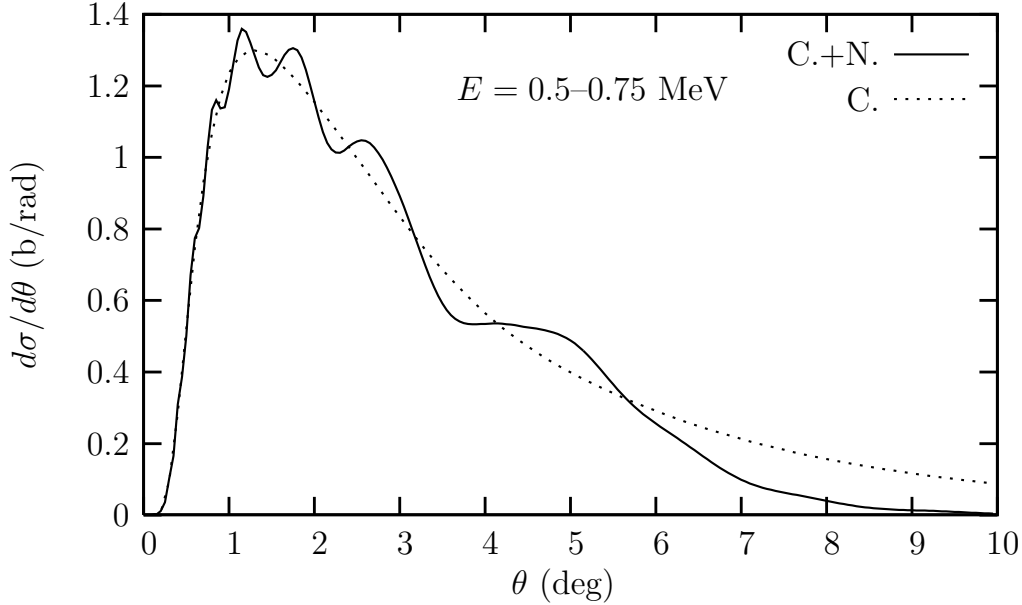}
\caption{Theoretical calculations of \ex{8}B breakup on Pb at 52~MeV/nucleon.
The angular distributions obtained for $E=0.5$--0.75~MeV
are not convoluted with the experimental resolution.
dynamical calculations are performed using
Coulomb plus nuclear (full lines) and purely Coulomb (dotted line)
$P$-$T$ interactions.}\label{f52nc}
\end{figure}

To complete this analysis of the influence of the nuclear interaction
as well as to show the effect of the filtering on the distributions,
we show in \fig{f52nc} the result of our calculations
for the first relative-energy bin before filtering.
The full line corresponds to the Coulomb plus nuclear $P$-$T$ interactions.
The dotted line is obtained with purely Coulomb interactions.
Both distributions exhibit the same magnitude for $\theta<6$\textdegr.
However, the former oscillates around the
latter. This effect is a signature
of the interferences between the solutions of \Eq{TDSE} corresponding
to neighboring impact parameters \cite{GBC06}.
The absence of such oscillations in the purely
Coulomb calculation shows that the presence of nuclear terms in the
$P$-$T$ potentials is responsible for this pattern,
already observed by Ogata \etal (see Fig.~2 of \Ref{Oga06}).
The angular distribution is thus sensitive to nuclear interactions
even at very forward angles.
However the experimental angular resolution hides
these interferences in the measurements.
Once filtered, both theoretical distributions look
very similar at forward angles (see \fig{f52}).
This result also explains that no effect of the nuclear interaction
is observed in the parallel-momentum and energy distributions.
In these cases, the integration over $\theta$ cancels out the oscillations.
The angular distribution seems therefore a proper observable
to bring out the influence of nuclear interactions.
However this requires a very fine experimental resolution.

To evaluate the contribution of the quadrupole term of the
Coulomb $P$-$T$ interaction to the angular distribution,
we also compute this observable
using a purely $E1$ interaction.
The corresponding calculations are displayed as
dashed lines in \fig{f52}.
As observed for the energy distribution (see \fig{f83}),
these results are not significantly different from the
purely Coulomb ones.
The difference between both calculations indeed remains small
in comparison with the size of the error bars and/or with
the influence of the nuclear interaction.
Moreover it varies with both the scattering angle and the relative energy.
The $E1$ distribution can indeed be higher or lower than
the total Coulomb one, confirming the energy dependence of the
$E1$-$E2$ interferences mentioned earlier.
Moreover, these results show the angular distribution to be an improper
observable to analyse the influence of the $E2$ strength
upon breakup, unlike the parallel momentum distribution.

We also compare the DEA with
perturbation theory. The dash-dotted lines in \fig{f52}
correspond to the first-order distribution obtained using
the classical relation between the impact parameter $b$ and the
scattering angle $\theta$.
As for the energy distribution, the first-order
approximation overestimates the dynamical calculation.
This is particularly true at large scattering angle
(i.e. small $b$ in a semiclassical point of view),
where perturbation theory is less valid.
However even below 1\textdegr, the difference between
the first-order and dynamical calculations remains of the order of 10\%.
This confirms the significant influence of higher-order effects
in the dissociation process, and the difficulty of interpreting experimental
data within the framework of first-order perturbation theory.

This analysis of the angular distributions indicates that
like parallel-momentum distributions, this observable
is relatively well described by our dynamical calculation
using a rather simple description of the projectile.
Angular distributions do not seem very sensitive to the
projectile description.
This result is confirmed by the similarity between our results
and those of Ogata \etal, which are obtained with different
models of \ex{8}B \cite{Oga06}.
Our analysis also shows that contrary to what was supposed by
Kikuchi \emph{et al}, this observable is not properly suited
to extract the $E2$ contribution to the breakup.
First there does not seem to be a range in angle where
the breakup is $E2$ dominated.
Second, as for the other observables, higher-oder effects play a significant
role, which makes hypothetical the use of
perturbation theory to analyse the data.
On the contrary, it seems that angular distributions
seem to be well suited to emphasize the effects of nuclear
interaction in dissociation.
An interference pattern indeed appears when optical potentials
are included in the $P$-$T$ interaction.
Unfortunately, the detection of this pattern requires
too fine a resolution to be observed experimentally in the
available measurements.

\subsection{On the extraction of the \ex{7}Be$(p,\gamma)$\ex{8}B 
astrophysical $S$ factor}\label{S17}
All the aforementioned Coulomb breakup experiments
have been performed with the final aim of determining
the astrophysical $S_{17}$ factor for the radiative
capture \ex{7}Be$(p,\gamma)$\ex{8}B at stellar energies.
The analyses presented in the preceding sections show that
the task is more complicated than initially suggested by
Baur \etal \cite{BBR86}.
The breakup process cannot be seen in these experiments
as a mere one step $E1$
transition from the initial bound state to the continuum.
If indeed the nuclear interaction between the projectile
and the target can be suppressed by selecting data at
forward angle, the presence of significant $E2$ strengths,
and higher-order effects hinder the direct extraction of
$B(E1)$ from the breakup data.

To circumvent these problems, one could think of using these measurements
to constrain the \ex{8}B description.
However, our systematic analysis of various experiments
shows that a crude description of \ex{8}B is sufficient
to explain most of the measurements.
The mechanism of Coulomb breakup seems therefore well understood.
However, this result also suggests that this reaction is not a very
accurate probe of the structure of the projectile.
A more realistic description of the projectile might
not improve significantly the agreement with experiment,
and therefore could hardly be constrained by such data.

As a first attempt to analyse the influence of the projectile
description upon the breakup calculations, we have computed
the energy distributions using three \ex{7}Be-$p$ potentials
adjusted in the $s$ wave to reproduce various scattering lengths
$a_0=5.9$~fm, $-2.8$~fm, and $-7$~fm (see \Sec{dE}).
Unfortunately, none of the descriptions seems to better fit the data.
We have observed a change only in the magnitude of the
cross section (see \fig{f83a0}).
If information about the \ex{8}B description were to be
extracted from this analysis, a different normalisation factor
would then be obtained for each potential.

\begin{figure}
\includegraphics[width=14cm]{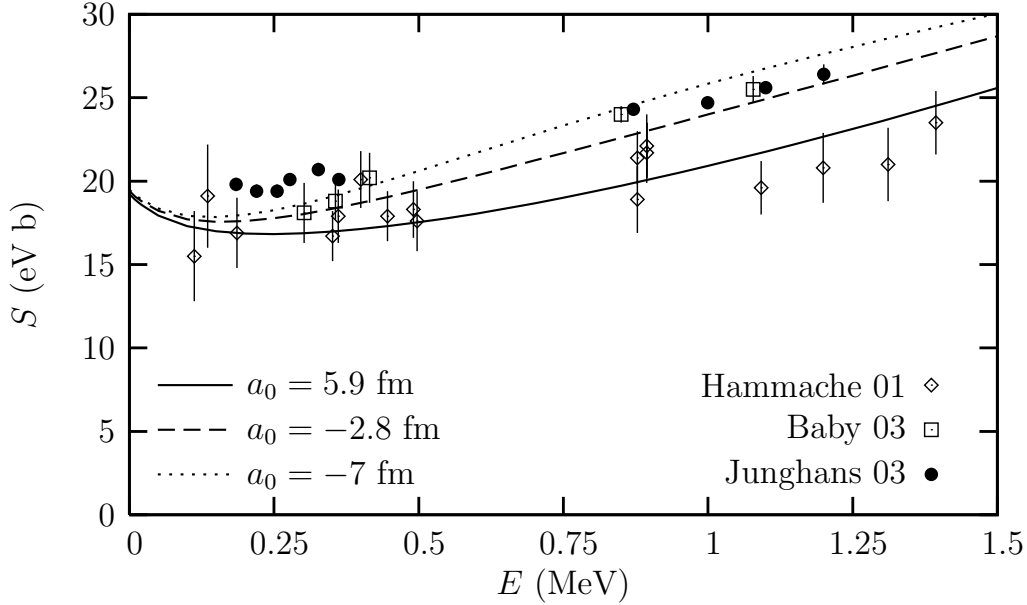}
\caption{Astrophysical $S_{17}$ factor for the
\ex{7}Be$(p,\gamma)$\ex{8}B radiative capture.
Calculations are performed with the three \ex{8}B descriptions
given in \Sec{dE}, characterized by the scattering lengths
$a_0=5.9$~fm (initial potential; full lines),
$a_0=-2.8$~fm (dashed lines), and $a_0=-7$~fm (dotted lines).
Experimental data are from Refs.~\cite{Ham01,Bab03l,Jun03}.}\label{fS17}
\end{figure}

To evaluate the influence of the \ex{8}B model
upon the radiative capture, we plot
the $S_{17}$ factor corresponding to
each of the three potentials in \fig{fS17}
as a function of the \ex{7}Be-$p$ relative energy $E$.
The results of three recent direct measurements are shown
as well \cite{Ham01,Bab03l,Jun03}.
All three potentials lead to very similar $S_{17}$ factors at zero energy:
19.2, 19.4, and 19.5~eV~b, for $a_0=5.9$~fm (full line),
$-2.8$~fm (dashed line), and $-7$~fm (dotted line) respectively.
This is due to the tiny dependence of the astrophysical $S_{17}$
factor on the scattering length at low energy \cite{Bay00}.
The energy dependence of $S_{17}$, however, differs from one
potential to the other. While it behaves rather smoothly
with $E$ in the $a_0=5.9$~fm case, it increases
more rapidly when $a_0$ is negative.
This smooth behavior is in agreement with the
measurements of Hammache \etal \cite{Ham01}, and Junghans \etal \cite{Jun03},
although both sets of data differ in magnitude.
The data of Baby \etal \cite{Bab03l} seem better reproduced
when $S_{17}$ increases faster with the energy.

These results show that there remain significant
uncertainties in the extraction of $S_{17}$
at stellar energy from breakup data.
In this indirect technique, the comparison between
theory and experiment provides information at two
levels.
First it evaluates the ability of both the reaction
and projectile-structure models to describe the
breakup mechanism, i.e. to reproduce accurately
the shape of the cross sections.
Second, it gives a scaling factor from the adjustment
of the magnitude of the theoretical cross section
upon the experimental one.
The $S_{17}$ factor extracted from breakup would
then be the theoretical $S_{17}$ obtained from
the \ex{8}B model multiplied by this scaling factor.
Our analysis shows that within the DEA, a crude
two-body description of \ex{8}B is able to reproduce
fairly well the shape of most of the Coulomb breakup
observables.
However, as seen in \Sec{dE}, different \ex{8}B
models, while leading to similar energy
distributions, give different magnitudes of this
cross section.
Since these models lead to essentially the same theoretical
$S_{17}$ at zero energy (see \fig{fS17}), one would
extract, from that observable, a different $S_{17}$
from each of these \ex{7}Be-$p$ potentials.
We also see that even with one description of
\ex{8}B, different $S_{17}$ could be extracted from
Coulomb breakup data, depending on the observable
considered.
Indeed, the good agreement between theory and
experiment for the parallel-momentum distribution
at 44~MeV/nucleon (see \fig{f44}) suggests a unit
scaling factor. However, the DEA calculation
performed with the same \ex{8}B description
overestimates the experimental parallel-momentum
distribution at 81~MeV/nucleon by up to 10--25\%
(see \fig{f81}), suggesting a scaling factor
lower than one, and therefore a smaller $S_{17}$
at zero energy.

The mechanism of the Coulomb breakup of \ex{8}B seems now well understood.
Most of the breakup observables are indeed well reproduced
with the DEA considering a simple two-body model of the projectile.
Nevertheless, efforts still need to be done to understand
the discrepancies between theory and experiment
in the magnitude of some cross sections.
Since this magnitude is sensitive to the \ex{7}Be-$p$ scattering length,
similar calculations involving a more realistic \ex{8}B model
(i.e. reproducing the measured scattering lengths, and therefore
including the spin of the core) might solve this problem.
The approximate treatment of special relativity used in the present
breakup model (see \Sec{rel}) may also be an issue.
To evaluate this effect, a comparison with a relativistic extension
of the DEA is needed. Such an extension is planned in the near future.
However, the recent results obtained by Bertulani with a relativistic
extension of the CDCC technique do not suggest this effect to be
significant at intermediate energies \cite{Ber05}.
Until then the extraction of the astrophysical $S_{17}$ factor
from breakup data will be subjected to uncertainty no smaller than
that obtained from direct measurements.

\section{Conclusion}\label{conclusion}
The cross section of the radiative capture reaction
\ex{7}Be($p,\gamma$)\ex{8}B is one of the key inputs to calculate
accurately the flux of high-energy neutrinos produced in the sun.
The difficulty of its direct measurement has raised interests in
indirect techniques. It has been proposed to infer the astrophysical
$S_{17}$ factor at stellar energies from
cross sections of the Coulomb breakup of \ex{8}B \cite{BBR86}.
Several experiments have been performed with this aim
\cite{Mot94,Kik97,Dav98,Gui00,Dav01l,Sch06}.

Unfortunately, the extraction of $S_{17}$ from the breakup of \ex{8}B
is not straightforward. While radiative capture is purely $E1$ dominated, 
breakup includes a non-negligible $E2$ strength \cite{EB96,Dav98}.
Moreover higher-order effects in the Coulomb breakup process
may hinder this extraction \cite{EB96}.
Finally, the projectile-target nuclear interaction,
albeit small, might spoil the dissociation data.

In this paper, we analyse various \ex{8}B Coulomb breakup sets of data
\cite{Dav98,Dav01l,Kik97} within a single reaction model
(the dynamical eikonal approximation \cite{BCG05,GBC06})
considering a single description of \ex{8}B.
This description is based on a simple \ex{7}Be-p potential model,
in which the spin of the core is neglected.
Although very crude, this description seems sufficient
to reproduce most of the breakup observables, computed within the DEA.
We indeed obtain a good agreement with experiment for angular and
longitudinal momentum distributions in both shape and magnitude.
Albeit fair, the agreement is less good for the energy distribution,
possibly because of an insufficiently realistic \ex{8}B description.

To analyse the breakup mechanism and its sensitivity to projectile-target
interactions, we systematically
perform all calculations with various projectile-target potentials.

First the calculations are done with optical potentials that simulate
both Coulomb and nuclear interactions. To analyse the influence
of the nuclear interaction on the data, they are compared to calculations
performed with purely Coulomb potentials.
In most of the cases the difference is found negligible if distributions
are limited to very forward angles.
The influence of nuclear interactions can thus be
avoided in Coulomb breakup.

Second, we compare the breakup observables obtained with either the full
Coulomb interaction or only its dipole term.
For all cross sections, we observe differences, but
only for the parallel-momentum distribution are they significant.
In that distribution, the $E1$ calculation exhibits a reverse
asymmetry compared to the full Coulomb one, confirming
the sensitivity of that observable to the $E2$ strength
revealed in previous works \cite{EB96,Dav98,Dav01c,MTT02}.

Finally, with the aim of analysing the significance of higher-order
effects, we compare our dynamical calculation with first-order
perturbation theory \cite{AW75}.
In agreement with previous analyses \cite{CB05,EBS05},
we observe significant couplings inside the continuum
leading to interferences between $E1$ and $E2$ first-order
transitions.
These interferences lead to a reduction of the asymmetry
of the parallel-momentum distributions, as observed in
Refs.~\cite{EB96,Dav98,Dav01c,MTT02,DT03}.

It seems therefore difficult to infer the accuracy
of the astrophysical $S_{17}$
factor extracted from breakup measurements as suggested in \Ref{BBR86}.
If indeed the projectile-target nuclear interaction can be neglected
by selecting forward scattering angles,
the non-negligible $E2$ contribution, and the presence of significant
higher-order effects limit the reliability of this extraction.
An unresolved issue is the influence of the \ex{8}B description
upon breakup calculations. The currently available data
do not seem to suggest such a significant effect.
But, perhaps the limited experimental acceptance
hides some interesting effects.
Therefore, future works are planned to evaluate the
interplay between the structure of \ex{8}B and its dissociation.
This requires an improvement of the projectile description
within the DEA.
If breakup turned out to be a useful probe of the structure of \ex{8}B,
it could serve to constrain a more precise \ex{8}B model.
Subsequently, this model could provide a reliable
extrapolation of $S_{17}$ down to low energies.

\begin{acknowledgments}
This text presents research results of the Belgian program P6/23 on
interuniversity attraction poles initiated by the Belgian-state
Federal Services for Scientific, Technical and Cultural Affairs (FSTC).
G. G. acknowledges the support of the FRIA, Belgium.
P. C. acknowledges the support of the Fund for Scientific Research (F.
R. S.-FNRS), Belgium.
\end{acknowledgments}

\appendix
\section{First-order perturbation theory}
\label{appendixA}
By using the first-order perturbation theory to solve the time-dependent
Schr\"odinger equation the breakup amplitudes used in \Eq{TDE}
are given by \cite{AW75}
\beq
	S^{(m_0)E\lambda}_{kljm}(b) &=& \frac{1}{i\hbar} Z_T Z_{\rm{eff}}^{(\lambda)} \frac{e^2}{4\pi\epsilon_0} e^{i(\sigma_l + \delta_{lj} - l\pi/2)} (-1)^{I-m_0} \sqrt{\frac{4\pi}{2\lambda+1}} I_{\lambda\ m_0-m}\nonumber \\
& & \times \sqrt{(2l_0+1) (2j_0+1) (2l+1) (2j+1)}
\left(\begin{array}{c c c} l & \lambda & l_0 \\ 0 & 0 & 0 \end{array}\right)
\left\{\begin{array}{c c c} j & l & I \\ l_0 & j_0 & \lambda \end{array}\right\} \nonumber \\
& & \times
\left(\begin{array}{c c c} j_0 & \lambda & j \\ m_0 & m-m_0 & -m \end{array}\right)
 \int_0^\infty u_{klj}(r) r^\lambda u_{n_0l_0j_0}(r)dr,
\eeqn{SkljmE1E2}
where $\omega=(E-E_0)/\hbar$, and the effective charge
$Z_{\rm{eff}}^{(\lambda)}$ is defined by
\beq
Z^{(\lambda)}_{\rm eff} = \left(-\frac{m_c}{m_P}\right)^\lambda Z_f + \left(\frac{m_f}{m_P}\right)^\lambda Z_c.
\eeqn{Zeff}
For straight line trajectories, the time integral $I_{\lambda \mu}$
\{see e.g. Eq.~(13) of \Ref{CB05}\}
in \Eq{SkljmE1E2} can be evaluated analytically \cite{EB02}
\beq
I_{\lambda \mu}&=& \sqrt{\frac{2\lambda+1}{4\pi}}\frac{2}{v} \frac{i^{\lambda+\mu}}{\sqrt{(\lambda+\mu)!(\lambda-\mu)!}} \left(\frac{\omega}{v}\right)^\lambda K_{|\mu|}\left(\frac{\omega b}{v}\right),
\eeqn{It}
where $K_n$ is a modified
Bessel function \cite{AS70}.
\section{Frame transformation}
\label{appendixB}
The core and fragment momentum in the laboratory frame are defined in spherical coordinates by
$\ve{p_c}=(p_c,\theta_c,\varphi_c)$ and $\ve{p_f}=(p_f,\theta_f,\varphi_f)$
while the total momentum $\ve{p}_{tot}$ is assumed to be in the $Z$-direction.
The energy and momentum conservation laws lead to the following relation
\beq
		& & p_f^2 \left( \frac{1}{2m_f} + \frac{1}{2m_T} \right)
		- \frac{p_f}{m_T} \left( P_{tot} \cos\theta_f - p_c \sin\theta_f\sin\theta_c\cos\Delta\varphi
		- p_c \cos\theta_f\cos\theta_c \right) \nonumber \\
	  & & + \left[ \frac{p_c^2}{2m_c} - E_{tot} + \frac{1}{2m_T} (p_{tot}^2 + p_c^2 - 2 p_{tot} p_c \cos\theta_c)  \right] = 0,
\eeqn{B1}
where $E_{tot}$ is the total energy in the laboratory frame. The fragment momentum $p_f$ can thus be deduced from $p_c,\theta_c,\theta_f$ and $\Delta\varphi=\varphi_c-\varphi_f$.
>From Eqs.~(\ref{CV1}) and (\ref{CV2}), one obtains in the center of mass frame, the relative momentum between the fragment and the core
\beq
	\hbar^2 k^2 = \left(\frac{m_c}{m_P}\right)^2 p_f^2
	        +\left(\frac{m_f}{m_P}\right)^2 p_c^2 
	     -2 \frac{m_f m_c}{m_P^2} p_f p_c \left( \sin\theta_f\sin\theta_c\cos\Delta\varphi + \cos\theta_f\cos\theta_c\right),
\eeqn{B2}
the corresponding colatitude
\beq
	\cos\theta_k = \frac{m_c p_f \cos\theta_f - m_f p_c \cos\theta_c}{m_P \hbar k},
\eeqn{B3}
the relative momentum between the projectile and the target
\beq
	\hbar^2 K'^2 &=& p_c^2 + p_f^2 + 2 p_c p_f \sin\theta_c \sin\theta_f \cos\Delta\varphi
				  + 2 p_c p_f \cos\theta_c \cos\theta_f \nonumber \\
			&	& + \left( \frac{m_P}{m_T+m_P} \right)^2 p_{tot}^2
			    - 2 \frac{m_P}{m_T+m_P} p_{tot} (p_c \cos\theta_c + p_f \cos\theta_f),
\eeqn{B4}
the corresponding colatitude
\beq
	\cos\theta = \frac{p_c \cos\theta_c + p_f \cos\theta_f- \frac{m_P}{m_T+m_P} p_{tot}}{\hbar K'},
\eeqn{B5}
and the difference between the two azimuthal angles
\beq
	\tan(\varphi-\varphi_k) = \frac{\sin\Delta\varphi}{\frac{m_c}{m_P} \frac{p_f \sin\theta_f}{p_c \sin\theta_c} - \frac{m_f}{m_P} \frac{p_c \sin\theta_c}{p_f \sin\theta_f} + \frac{m_c-m_f}{m_P} \cos\Delta\varphi}.
\eeqn{B6}
The three-body phase space factor is given by \cite{Fuc82}
\beq
	\rho (E_c,\Omega_c,\Omega_f) &=& \frac{m_c m_f m_T p_c p_f}{(2\pi\hbar)^6} \Big[ m_T + m_f + \frac{m_f}{p_f} (p_c 				
				\sin\theta_f\sin\theta_c\cos\Delta\varphi  \nonumber \\
				& &  + p_c \cos\theta_f\cos\theta_c - p_{tot} \cos\theta_f) \Big]^{-1}.
\eeqn{B7}

%\bibliography{abbrev,mybiblio,biblio,misc}

\begin{thebibliography}{10}
\bibitem{B8bdp}
 Note the typo in Eq.~(45) of \Ref{GBC06}, where the factor
  $i^{|m-m_0|}$ is wrongly placed at the numerator. The correct expression of
  the breakup transition matrix element is given by \Eq{TDE}.
\end{thebibliography}
%\bibliographystyle{prsty}

\end{document}